\documentclass[12pt]{article}

\usepackage{color}
\usepackage{enumitem}
\usepackage{times}
\newenvironment{sciabstract}{%
\begin{quote} \bf}
{\end{quote}}
\usepackage{graphicx}  
\usepackage{cite}
\usepackage{amsmath,amssymb,amsfonts}
\usepackage{amsthm}

\newtheorem{thm}{Theorem}

\newcommand{\bra}[1]{\langle{#1}\vert}
\newcommand{\ket}[1]{\vert{#1}\rangle}
\newcounter{lastnote}

\title{Provable Quantum Advantage in Randomness Processing}
\author
{Howard Dale, David Jennings and Terry Rudolph\\
\\
\normalsize{Department of Physics, Imperial College London}\\
\normalsize{Prince Consort Road, London SW7 2AZ, United Kingdom}\\
}

\date{}

\begin{document} 
\maketitle

\begin{sciabstract}
Quantum advantage is notoriously hard to find and even harder to prove. For example the class of functions computable with classical physics actually exactly coincides with the class computable quantum-mechanically. It is strongly believed, but not proven, that quantum computing provides exponential speed-up for a range of problems, such as factoring. Here we address a computational scenario of ``randomness processing'' in which quantum theory provably yields, not only resource reduction over classical stochastic physics, but a strictly larger class of problems which can be solved. Beyond new foundational insights into the nature and malleability of randomness, and the distinction between quantum and classical information, these results also offer the potential of developing classically intractable simulations with currently accessible quantum technologies.

\end{sciabstract}

Suppose you are handed a classical coin with some \emph{unknown} bias, is there a method by which one can simulate a perfectly fair coin-flip? The folklore method (attributed to von-Neumann [1]) is as follows: Flip the coin twice, if the two outcomes are different then output the coin with its value on the second flip, otherwise try again. Provided that the unknown probability of  heads, $p$ is not $0$ or $1$, it is clear that this method yields, with probability one, an unbiased output after a random number of coin-flips. Contrast this with the case where one is asked to output a new coin that has probability of heads $p^2$. In this case exactly two flips suffice for any $p\in[0,1]$: we flip the biased coin twice and if both flips are heads then we output the new coin showing heads, otherwise we output it showing tails. 

These two examples tell us that the output bias functions $f(p):=1/2$ and $f(p):=p^2$ can both be ``constructed'' by flipping a coin with some unknown bias $p$.
More generally, we say that a function $f$ is \emph{constructible}\footnote{In the mathematics literature the words 
\emph{observable} and \emph{simulable} are often used, however these already have unrelated but 
potentially confusing technical meaning within quantum theory.} if there is an algorithm that yields an output with bias $f(p)$, almost surely after a finite number of coin-flips. Slightly more precisely: for all $p$ the procedure to construct $f(p)$ must define disjoint sets $S_{1}$ and $S_{2}$ whose elements are finite strings, such that with probability $1$ the sequence of flips produced has exactly one of these strings as an initial segment. For any sequence of flips the output is heads if the initial segment is in $S_{1}$ and tails if it is in $S_{2}$, and thus an output is produced almost surely in finite time.

The topic of which functions are constructible, how easily they can be constructed, and their applications goes by the name ``Bernoulli Factory'' [2-5]. Crucially, in 1995 a theorem of Keane and O'Brien [2] determined the exact set of functions constructible from a classical coin of unknown bias $p$. Loosely speaking, it was found that a function $f(p):(S\subseteq[0,1])\rightarrow [0,1 ]$ is constructible if and only if (a) it is continuous, (b) it does not touch $0$ or $1$ within its domain and (c) it does not approach zero or one exponentially quickly at any edge of its domain (see the supplementary materials for a precise statement). While this allows such surprising functions as $e^{\cos p}$, $\sqrt{p}$, it also rules out important ones such as as the ``probability amplification'' function $f(p)= 2p$, which is central to certain stochastic simulation protocols. Moreover, it says nothing about the resources required to actually construct the functions -- often an infeasibly large number of coin-flips are required.

The scenario described for the Bernoulli factory shares similarities to a Turing Machine, however it is worth emphasizing that there are differences between it and a fully universal scenario. While both possess a target function to be ``computed'', the Bernoulli factory, with its unbounded, probabilistic i.i.d input, is in a sense a simpler, arguably more tractable, model. This makes the Bernoulli Factory an ideal candidate with which to establish quantum-mechanical results that are provably beyond the reach of classical physics. 

The central results we present in this work are:
\begin{enumerate}
\item Quantum Bernoulli Factories allow the construction of a strictly larger class of functions than allowed in stochastic classical physics.
\item Quantum Bernoulli Factories provide dramatic improvements in terms of resource requirements over a range of classically constructible functions.
\end{enumerate}

The classical Bernoulli Factory (CBF)  can be easily described within a quantum-mechanical setting via (arbitrarily many) copies of a qubit prepared in the mixed qubit state 
\begin{equation}\rho=p|0\rangle\langle0|+(1-p)|1\rangle\langle1|,\end{equation} for unknown 
$p\in S\subseteq[0,1]$. Here the computational basis $\{|0\rangle,|1\rangle\}$ of the two-dimensional qubit Hilbert space $\mathcal{H}$, denotes a fiducial projective measurement that extracts classical data. In contrast, a quantum-mechanical extension of the classical coin states has coherences in this basis. Our goal is to contrast the fundamental processing of such classical randomness with the quantum randomness attainable in a quantum Bernoulli factory (QBF). It should be emphasised, however, that our desired output is still classical. We refer to the quantum-mechanical extension of the coin state as a \emph{quoin}, and accordingly it is described by the coherent state 
\begin{equation}
|p\rangle\equiv\sqrt{p}|0\rangle+\sqrt{1-p}|1\rangle.
\end{equation}
For this we find measurement in the computational basis $\{|0\rangle,|1\rangle\}$ returns the probability distribution $(p, 1-p)$, and so by restricting to stochastic mixing in this basis, together with algorithmic processing, we see that the classical Bernoulli Factory setting is recovered as a special case from the quantum-mechanical one. In particular we see that the stochastic mixing available in the classical factory is a special case of the unitary operations available in the quantum setting\footnote{One could consider convex combinations of unitaries, however this turns out to be equivalent, since it is always possible to simulate this stochasticity via unitary generation of randomness on a subset of quoins or ancillae qubits.}.

\begin{figure}[H]
\begin{center}
\includegraphics[width=0.25\linewidth]{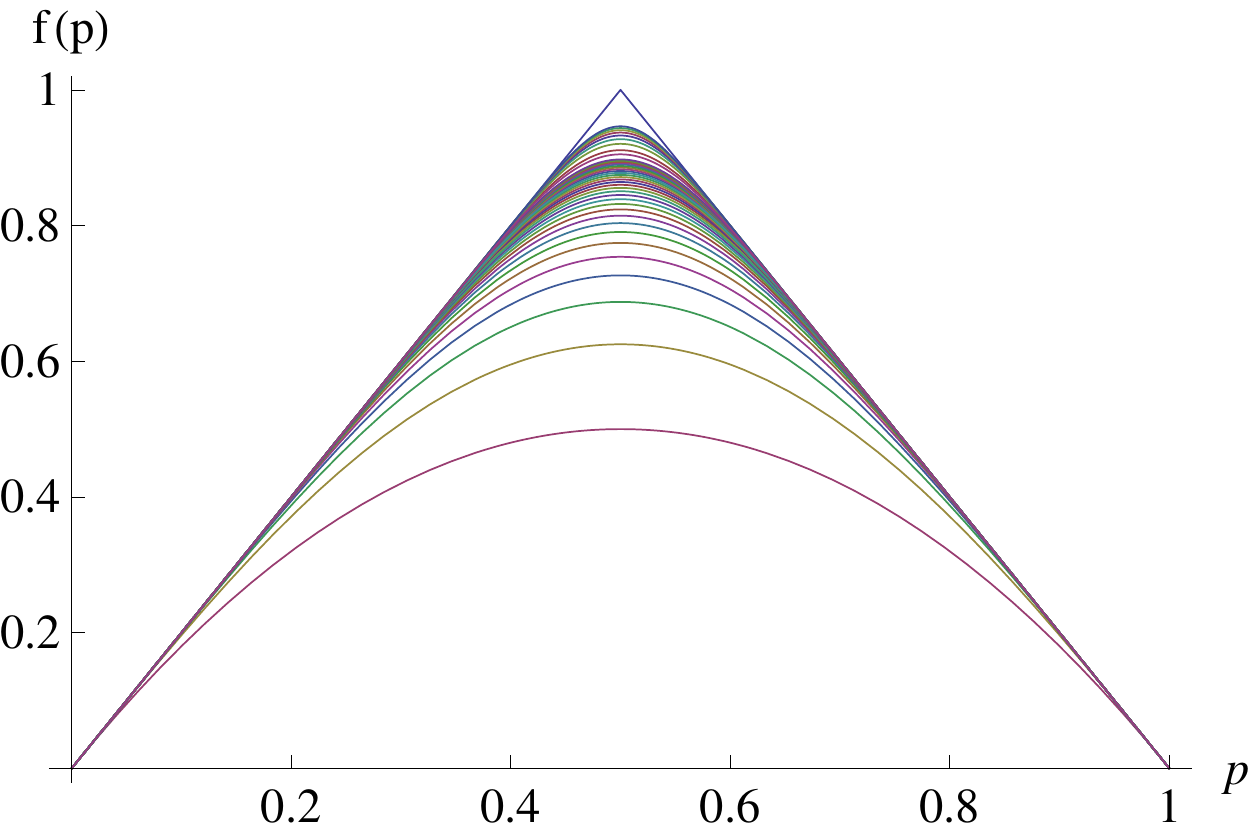}
\caption{\label{2pfig}The probability amplification function $f_\wedge: [0,1] \rightarrow [0,1]$, and the first 35 functions in the convex decomposition (\ref{2p}). This function is impossible to construct classically, but can be achieved quantum-mechanically via two-qubit measurements in the entangled Bell-basis, together with classical processing.}
\end{center}
\end{figure}

We now demonstrate that the full set of quantum-mechanical operations allows a strictly larger class of functions than allowed classically. A crucial demonstration of the superior power of the quantum-mechanical Bernoulli Factory is given by the probability amplification function $f(p)=2p$. This function is impossible to construct classically, since it attains the value $f(p)=1$ for $p=1/2$, and so traditional work-arounds involve ``chopping'' the function as it approaches $p=1/2$, and forming a truncated function, $f(p)= \min(2p,1-\epsilon)$  for some fixed $0<\epsilon <1$. This approximate function then does satisfy the conditions of the Keane O'Brien theorem, however the amount of coins needed to produce such a function scale very poorly with $\epsilon$ (see [6,7] for examples). In contrast we show that within a QBF it is possible to efficiently construct the classically impossible probability amplification function $f_\wedge:[0,1]\rightarrow[0,1]$ defined by:
\begin{displaymath}
   f_\wedge(p) := \left\{
     \begin{array}{lcr}
       2p & ; &p \in [0,1/2]\\
       2(1-p) & ; &p \in (1/2,1]
     \end{array},
   \right.
\end{displaymath} 
as shown in Fig.\ref{2pfig}.

Our method is as follows. The target function admits an alternative representation of $f_\wedge(p)=1-\sqrt{1-4p(1-p)}$, which in turn possesses an expansion of the form
\begin{equation}\label{2p}
 f_\wedge(p)=\sum_{k=1}^{\infty}{2k \choose k}\frac{1}{(2k-1)2^{2k}}(4p(1-p))^{k}:=\sum_{k=1}^{\infty}q_{k}(4p(1-p))^{k},
\end{equation}
where $q_k$ is a probability distribution.

Since within the (classical or quantum) Bernoulli Factory we can generate any constant distribution, we first construct an integer output $k$ with probability
\begin{equation}
q_{k}:={2k \choose k}\frac{1}{(2k-1)2^{2k}},
\end{equation}
and then conditioned on this output construct the function $g_k (p) = (4p(1-p))^k$. The latter set of functions $\{g_k\}$ are classically inaccessible for all $k>0$. We also note that $g_k(p) = g_{1}^{k}(p)$ and so our task reduces to constructing the $k=1$ case.

This is easily achieved by considering a Bell-basis measurement 
\begin{equation}
\{\ket{\phi^\pm}=(\ket{00}\pm\ket{11})/\sqrt{2},\ket{\psi^\pm}=(\ket{01}\pm\ket{10})/\sqrt{2}\},
\end{equation}
on two quoins. The probability that we obtain $\ket{\psi^+}\bra{\psi^+}$ or $\ket{\phi^-}\bra{\phi^-}$ is $1/2$, however the probability of obtaining the outcome $\ket{\psi^+}\bra{\psi^+}$, \emph{conditioned} on obtaining $\ket{\psi^+}\bra{\psi^+}$ or $\ket{\phi^-}\bra{\phi^-}$ is exactly $4p(1-p) \equiv g_1(p)$. Putting everything together, to construct a $f_\wedge(p)$ coin we output an index $k$ with probability $q_k$ and then construct $k$ $g_1(p)$-coins using $\mathcal{O}(k)$ quoins. If $k$ outcomes of heads in a row are obtained from the $g_1(p)$-coins then heads is output, otherwise tails is output. This provides an exact construction of the function $f_\wedge$, as claimed.

\begin{figure}[htp]
\begin{center}
\includegraphics[scale=0.7]{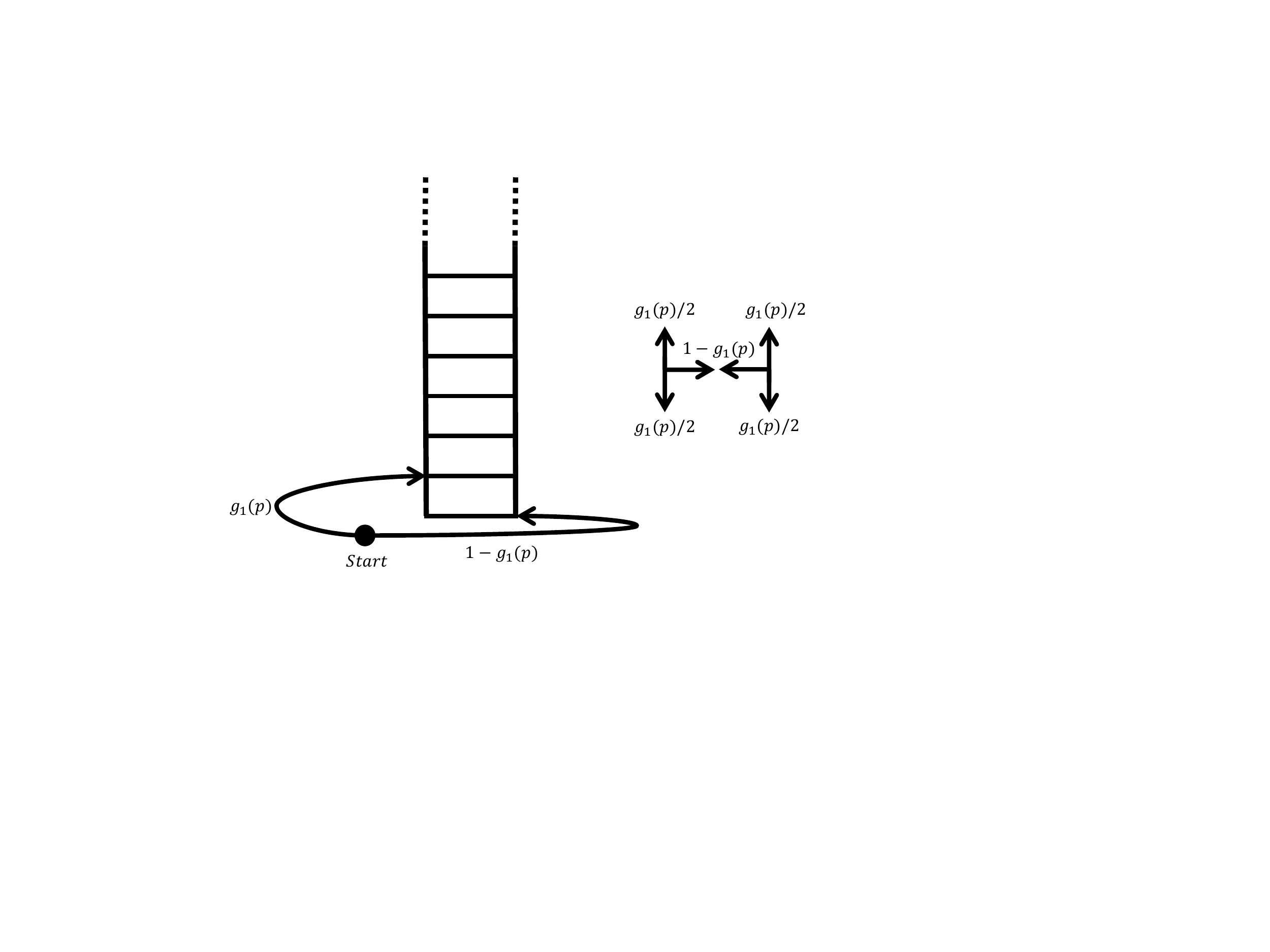}
\caption{\label{fig:walk} The quantum probability amplification construction can be represented via a random walk on a semi-infinite ladder with transitions given by Bell-measurement outcomes on two qubits. }
\end{center}
\end{figure}

We can provide a clearer account of this construction by adapting a method in [8], where we can represent the above method as a random walk on a ladder [Fig.\ref{fig:walk}]. One begins at the point marked ``Start" and flips a $g_1(p)$-coin to decide where to move next. Once on the ladder at any vertex we step up the ladder with probability $g_1(p)/2$, or down the ladder with the same probability; otherwise we move across. If we reach the bottom left corner we output heads, while if we reach the bottom right corner we output tails; this means that if the very first flip is tails we output tails immediately. The probability of outputting heads can be shown to be

\begin{align}
P(\text{heads})&=g_1(p)\frac{1-\sqrt{1-g_1(p)}}{g_1(p)}\\
&=1-\sqrt{1-g_1(p)}\\
&=1-\sqrt{1-4p(1-p)}=f_{\wedge}(p).
\end{align}

The construction that we have provided uses two-qubit measurements in an entangled basis, and so one might think that entanglement is required for any quantum advantage, surprisingly this is not the case -- the following theorems determine the exact class of functions $f:[0,1] \rightarrow [0,1]$ that are constructible within a QBF using only single-qubit operations, and are the main results of this work. In fact the only type of operations required for our proofs are the unitaries
\begin{equation}
H(a)=\left(\begin{array}{cc}\sqrt{1-a}&\sqrt{a}\\ \sqrt{a}&-\sqrt{1-a}\end{array}\right),
\end{equation}

which construct a coin that gives output one with probability

\begin{equation}
h_{a}(p)=\left(\sqrt{p (1-a)}-\sqrt{a(1-p)}\right)^{2}.
\end{equation}

\begin{thm}\label{Theorem1}
A function $f:[0,1]\rightarrow [0,1]$ is constructible with quoins and a finite set of single-qubit operations if and only if the following conditions hold.
\begin{enumerate}
\item $f$ is continuous.
\item Both $Z=\left\{ z_{i} : f\left(
z_{i}\right) =0 \right \}$  and $W=\left\{ w_{i} :f\left(
w_{i}\right) =1 \right \}$ are finite sets.
\item $\forall z\in Z$ there exists constants $c,\delta >0$ and and integer $k<\infty $ such that 
\begin{equation*}
c\left( p-z\right) ^{2k}\leq f\left( p\right) ~~\forall p\in \left[ z-\delta
,z+\delta \right]. 
\end{equation*}
\item  $\forall w\in W$ there exists constants $c,\delta >0$ and an integer $k<\infty $ such that 
\begin{equation*}
1-c\left( p-w\right) ^{2k}\geq f\left( p\right) ~~\forall p\in \left[
w-\delta ,w+\delta \right] .
\end{equation*}
\end{enumerate}
\end{thm}

The proof of Theorem 1 is too long to include here and is provided in the supplementary materials. The main idea is similar to the construction of the probability amplification function, and involves arriving at a convex decomposition in terms of functions that are explicitly constructible using quantum operations on quoins. It is clear that the conditions of the theorem are a natural generalization of the classical case, except now the function is allowed to go (polynomially quickly) to $0$ and $1$ at a finite number points over the interval $[0,1]$. Moreover, this implies that the scaling of resources within the interior no longer behaves as in the classical case, where large number of coins are required if the function approaches $0$ or $1$ at for example $p=1/2$. Instead, the scaling for any point $x \in [0,1]$ behaves like the end-points $p=0,1$.

One straightforward generalization is that we do not require the target function be defined at all points inside the interval $[0,1]$, and can allow more extreme behaviours (such as rapidly increasing oscillations or sharp discontinuities) in the functions that we construct. To this end we have the following theorem.
\begin{thm}\label{Theorem2}
A function $f:(0,a_{1})\cup(a_{1},a_{2})\cup ...\cup(a_{n},1)\rightarrow [0,1]$ is constructible with quoins and a finite set of single qubit unitaries if $f$ is continuous on its domain and there exists a finite list $\lbrace a_{1},a_{2},...a_{n'} \rbrace$, which contains $\lbrace a_{1},a_{2},...a_{n} \rbrace$, and integer $k$ such that 
\begin{equation}
a^{k}(p) \leq f(p) \leq 1-a^{k}(p)
\end{equation}
for all $p \in (0,1)$, where
\begin{equation}
a(p) := p(1-p)\prod_{1\leq i \leq n'} h_{a_{i}}(p)(1- h_{a_{i}}(p))
\end{equation}
\end{thm}

\begin{figure}[htp]
\begin{center}
\includegraphics[scale=0.6]{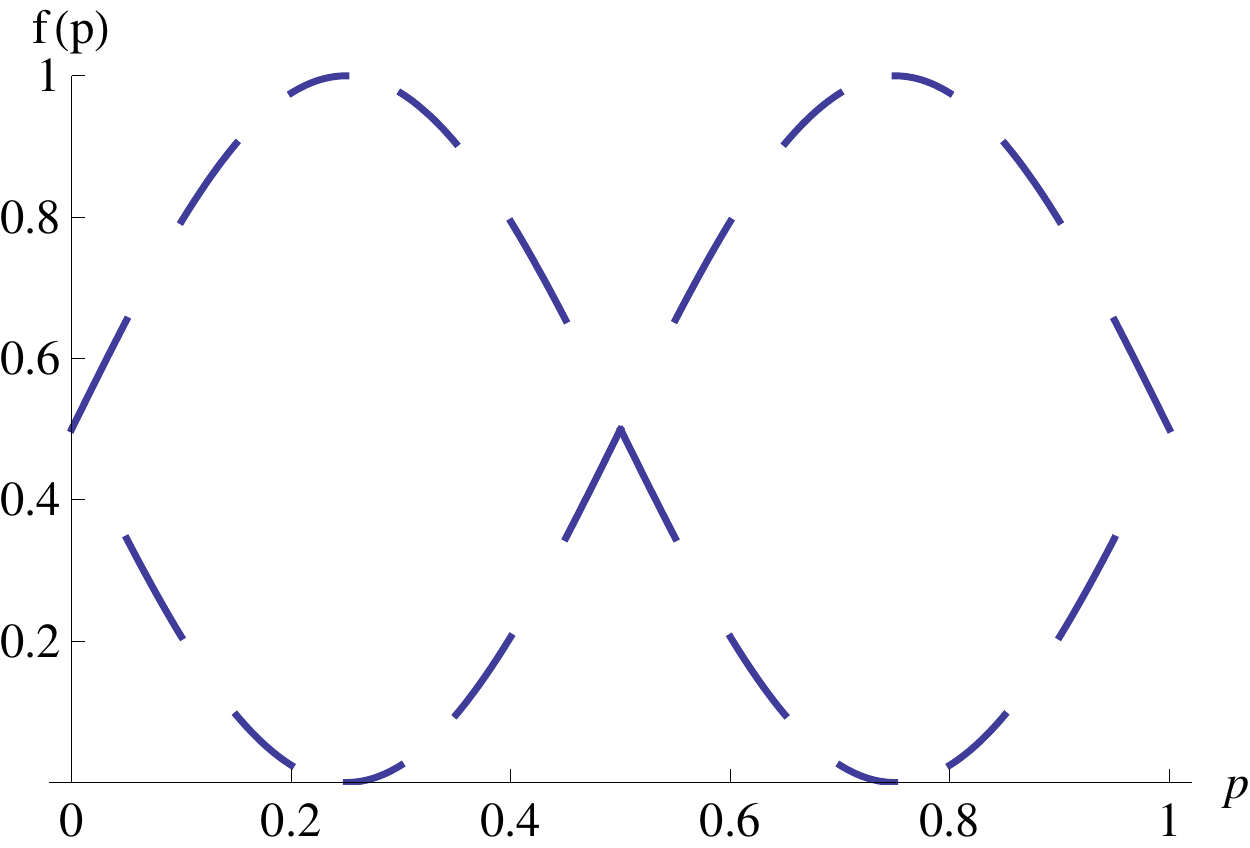}
\caption{\label{fig:wf} One example of the unusual functions one could construct using Theorem 2, the actual discontinuities are excluded. Effectively we gain access to piecewise continuous functions but with their discontinuous points excluded, leaving a function which is continuous on its domain.}
\end{center}
\end{figure}

The proof of Theorem \ref{Theorem2} follows a slightly different construction to Theorem \ref{Theorem1}, and is also provided in the supplementary materials. The above two theorems both relate to single-qubit operations and provide a broad class of constructible functions, however we conjecture that multi-qubit unitaries do not extend the set of quantumly constructible functions, but do provide additional speed-ups, as is illustrated by the example of the function $g_1$ constructed from Bell measurements. 

It is important to note that in addition to extending the class of functions, the quantum-mechanical Bernoulli Factory provides dramatic speed-ups for certain functions that are classically accessible. For example, consider the function $f_\alpha :[0,1] \rightarrow [0,1]$, given by $f_\alpha(p)=\alpha h_a(p)$ with $0<\alpha <1$, which is easily constructed via a convex combination of $h_a$ (which requires just a single quoin) and the function $0$. Since the function $h_a$ is inaccessible classically, the construction of the function $f_\alpha$ necessarily requires a rapidly increasing number of classical coins as $\alpha$ tends to $1$, in stark contrast to the quantum-mechanical case.

Up to this point we have been concerned with constructing from a coherent input a target distribution $f(p)$ as a classical probabilistic distribution, however a more sophisticated goal is to construct a \emph{coherent} output from a coherent input. Specifically, given an unbounded number of  input quoins $|p \rangle$ what output quoins $|f(p)\rangle$ can be obtained through arbitrary quantum operations on the input string? This is related to exact sampling tasks of the following form: given a classical algorithm that can efficiently sample bit strings $x_i$ with associated probabilities $p_i$, does there exist an efficient quantum algorithm which outputs the 
state
\begin{equation}
|\psi\rangle=\sum_k \sqrt{p_i} |x_i\rangle?
\end{equation}
This problem has been called \emph{q-sampling} [9,10] and q-samples constructible using the simple techniques of  [9-11]  form a useful starting point in many modern quantum algorithms [12-15] and enable q-sampling of various useful distributions [16-18]. It should be noted that the question of which efficient classical sampling algorithms allow for creation of an efficient q-sample is remarkably subtle.  For instance, it is classically trivial to uniformly sample the $n!$ adjacency matrices corresponding to permutations of the vertex labellings of an $n$-vertex graph, such a q-sample would easily allow for an efficient quantum algorithm to solve graph isomorphism but despite attempts from many researchers no efficient procedure has been found.

The distinguishing features of quantum and classical information are subtle, and often well-hidden. Paradigmatic examples have already appeared in single-party cryptography [19], two-party cryptography, and communication complexity [20]. Of arguably broader significance is to determine the computational abilities allowed by quantum physics. Quantum computing does not allow new functions to be constructed and the speed-ups, whilst strongly supported by evidence remain unproven. The work presented here provides a computational scenario in which quantum mechanics has strict superiority over classical physics and, by virtue of requiring only single-qubit manipulations, appears vastly easier to attain experimentally.

\section*{References}

\begin{enumerate}[label={[\arabic*]}]

\item J. Von Neumann. Various techniques used in connection with random digits. Applied Math Series, 12(36-38), 1951

\item  M. S. Keane and G. L. O'Brien. A Bernoulli factory. ACM Trans. Model. Comput. Simul. 4(2):213-219, April 1994

\item S. Asmussen, P. Glynn and H. Thorisson, Stationarity detection in the initial transient problem. ACM Transactions on Modeling and Computer Simulation 2(2):130-157, 1992

\item K. Latuszynski, I. Kosmidis, O. Papaspiliopoulos and G. O. Roberts. Simulating events of unknown probabilities via reverse time martingales. Random Structures and Algorithms, 38(4):441-452, 2011

\item J. Wastlund. Function arising by coin-flipping. Technical Report. KTH, Stockholm, 1999

\item A. C. Thomas and J. H. Blanchet. A Practical Implementation of the Bernoulli Factory. Preprint at http://arxiv.org/abs/1106.2508 (2011)

\item S. Nacu and Y. Peres. Fast simulation of new coins from old. The Annals of Applied Probability, 15(1A):93-115, 2005

\item] E. Mossel and Y. Peres. New coins from old: computing with unknown bias. Combinatorica, 25(6):707-724, 2005.

\item Lov Grover and Terry Rudolph. Creating superpositions that correspond to efficiently integrable probability distributions. \emph{arXiv preprint quant-ph/0208112, 2002}

\item Dorit Aharonov and Amnon Ta-Shma. Adiabatic quantum state generation and statistical zero knowledge.
In Proceedings or the thirty-fifth annual ACM symposium on Theory of computing, pages 20-29, ACM, 2003

\item S. Aaronson and A. Drucker, Advice Coins for Classical and Quantum Computation, Lecture Notes in Computer Science Volume 6755, 2011, pp 61-72

\item Chen-Fu Chiang, Daniel Nagaj, and Pawel Wocjan. 2010. Efficient circuits for quantum walks. Quantum Info. Comput. 10, 5 (May 2010), 420-434

\item A. W. Harrow, A. Hassidim and S. Lloyd. Quantum algorithm for linear systems of equations. PRL, 103(15):150502, 2009

\item I. Kassal, S. P. Jordan, P. J. Love, M. Mohseni and A. Aspuru-Guzik. Polynomial-time quantum algorithm for the simulation of chemical dynamics. Proceedings or the National Academy of Sciences. 105(48):18681-18686, 2008

\item K. Temme, T. J. Osborne, K. G. Vollbrecht, D. Poulin and F. Verstraete. Quantum metropolis sampling. Nature, 471(7336):87-90, 2011.

\item M. Jerrum and A. Sinclair. Approximating the permanent. SIAM journal on computing. 18(6):1149-1178, 1989

\item A. Frieze, R. Kannan, N. Polson et al. Sampling from log-concave distriubutions. The Annals of Applied Probability. 4(3):812-837, 1994

\item M. Dyer, A. Frieze and R. Kannan. A random polynomial-time algorithm for approximating the volume of convex bodies. Journal of the ACM. 38(1):1-17. 1991

\item P. W. Shor and J. Preskill. Simple proof of security of the bb84 quantum key distribution protocol. PRL 85(2):441, 2000

\item C. Brukner, M. Zukowski, Jian-Wei Pan and A. Zeilinger. Bell's inequalities and quantum communication complexity. PRL. 92(12):127901-127901, 2004

\end{enumerate}

\begin{figure}[htp]
\begin{center}
\includegraphics[scale=0.7]{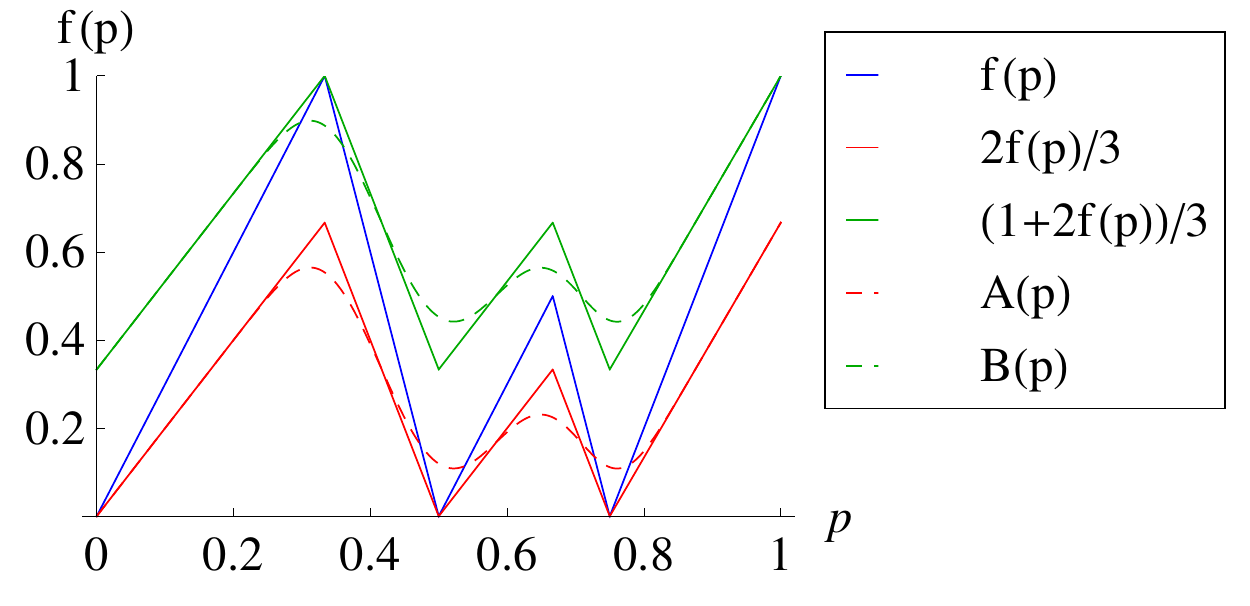}

\textbf{Supplementary Figure 1: Unsuccessful Bounding Functions} We attempt to bound the function from above and below. Here we show $A_{100}(p)$ and $B_{100}(p)$, and their counterparts for very large $n$, for the piecewise linear function depicted. Note that both fail to be good bounding functions when $f(p)$ gets close to zero or one.
\end{center}
\end{figure}

\begin{figure}[htp]
\begin{center}
\includegraphics[scale=0.7]{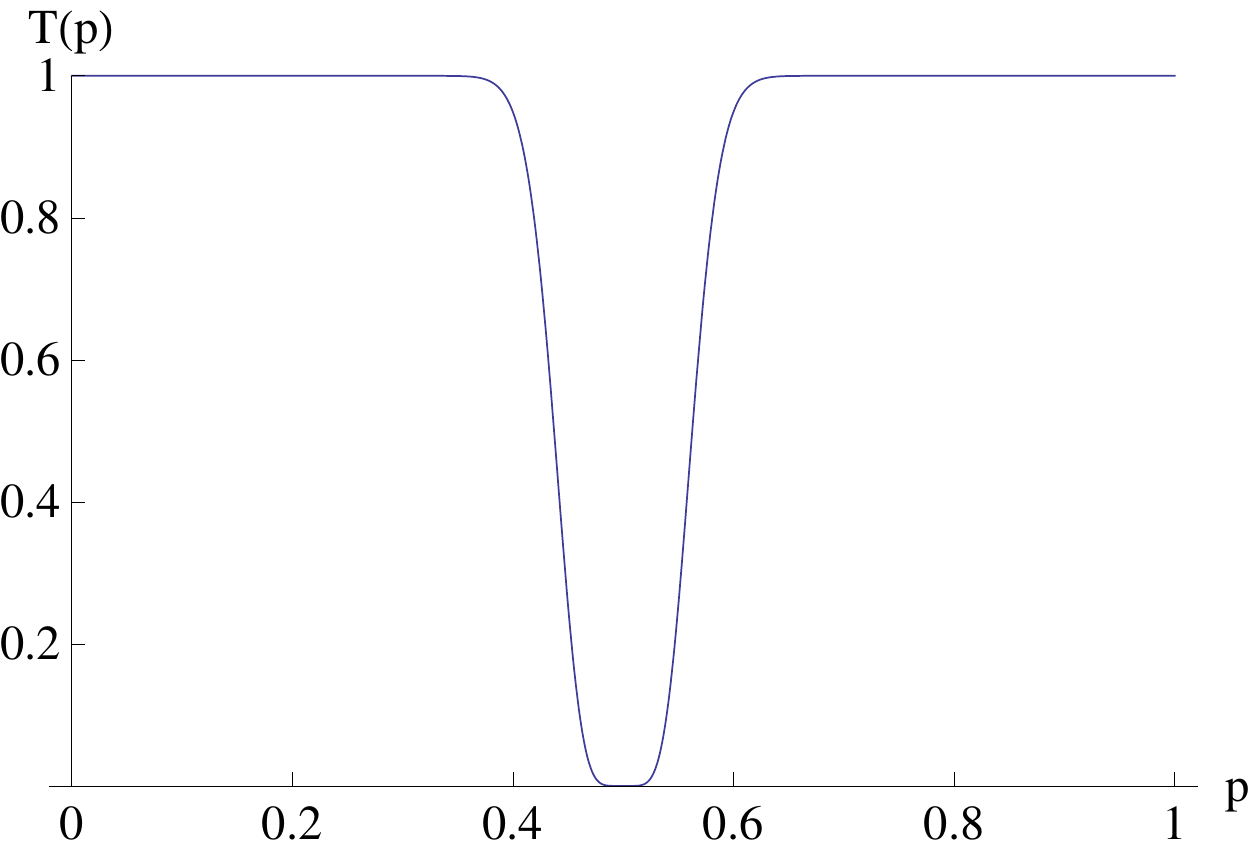}

\textbf{Supplementary Figure 2: Our q-constructible ``Heaviside" function} Here we show an example of an important function for this task, which approaches zero rapidly close to the chosen point. This function is $T_{400,3,0.5}$
\end{center}
\end{figure}

\begin{figure}[htp]
\begin{center}
\includegraphics[scale=0.7]{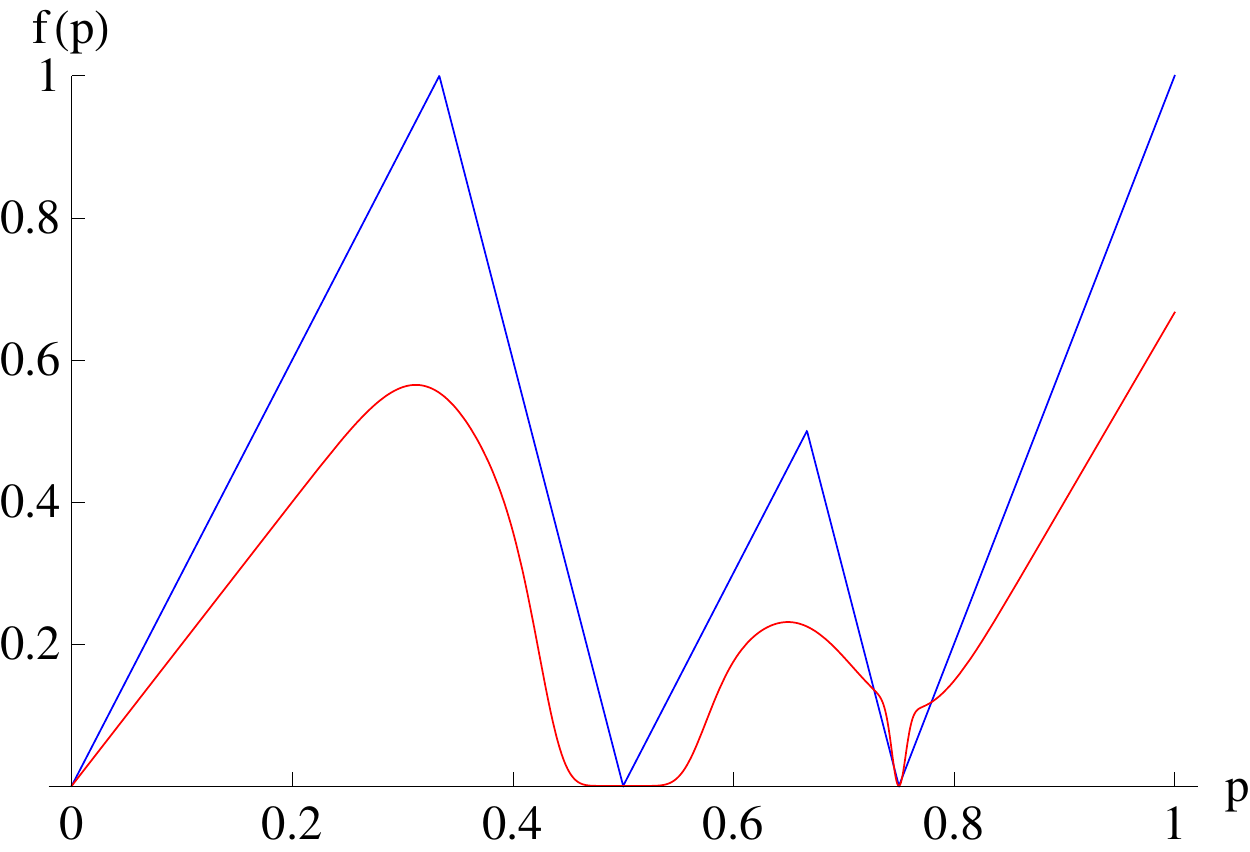}

\textbf{Supplementary Figure 3: Further Difficulties with Bounding Functions} The product function $A_{100}(p)T_{400,5,1/2}(p)T_{400,1,3/4}(p)$ is \emph{nearly} a good lower bounding function for $f(p)$. It struggles at $0.5$ from being pushed too flat and at $0.75$ from not being flat enough.
\end{center}
\end{figure}

\newpage
\newpage

\section*{Supplementary Methods}

In order to prove our two theorems we will need to define certain sets and probability measures.

We define $\Omega$ as the set of all infinite sequences $X=(X_{1},X_{2},X_{3},...)$ such that $X_{i} \in \lbrace 0,1 \rbrace$. A flip of ``heads" corresponds to the value $1$.

We define cylinder sets $V(v_{1},...,v_{k}) = \lbrace X \in \Omega : X_{i}=v_{i},1\leq i \leq k \rbrace$ as the set of all infinite sequences beginning with the string $(v_{1},...,v_{k})$. For example, $V(101)$, is the set of all infinite $\lbrace 0,1 \rbrace$ strings beginning with $101.$ The product topology on $\Omega$ is defined by taking these cylinder sets as a basis for the open sets.

A probability measure $P_{p}$ takes an open subset of $\Omega$ to the probability of it occurring. In the simplest case, that of classical coins with no processing, every value of $p\in[0,1]$, determines a probability measure $P_{p}$ on $\Omega$ such that $P_{p}(X_{i}=1)=p$ for every $i$.

For any subset $S$ of $\Omega$ we define the outer measure, required for when $S$ is not open, as

\begin{equation}
P_{p}(S)=\text{inf}\lbrace P_{p}(E):E \text{ is open and } S \subseteq E \rbrace,
\end{equation}

 An event is a subset of $\Omega$ which is measurable with respect to $P_{p}$ for every $p$.

The only events we are interested in discussing are those for which we can say whether they occur after some finite number of coin flips. We call these events discernible with respect to a given probability distribution. An event is discernible if the probability that we need to flip infinitely many coins to decide whether it occurs is zero. More formally:

\emph{Definition 1:} An event $S$ is discernible with respect to a probability measure $P_{p}$ if $P_{p}(S)+P_{p}(S^{c})=1$. This may be the case for some values of $p$ and not others.

Now in our quantum factory our ability to apply different unitaries to different coins means defining some new things.

The total sample space is given by
\begin{equation}
\Omega_{tot}=\Omega_{1}\times\Omega_{2}\times ...\times\Omega_{n}
\end{equation}
 is the set of all ordered $n$-tuples of infinite $\lbrace 0,1\rbrace$ strings. 

Subsets are then products of cylinder sets.
\begin{equation}
V_{tot}=V(\underline{v}_{1})\times V(\underline{v}_{2})\times ...\times V(\underline{v}_{n})
\end{equation}
is the set of all ordered $n$-tuples of infinite $\lbrace 0,1\rbrace$ strings which begin with the finite strings $\underline{v}_{1},\underline{v}_{2},...,\underline{v}_{n}$ respectively. Again illustrating with an example, the set $V(1)\times V(01)\times V(111)$ is the set of all ordered $3$-tuples of infinite $\lbrace 0,1\rbrace$ strings which begin with $1$, $01$ and $111$ respectively.

We also have different probability distributions in the quantum cases, defined by a value $p$ and finite set of unitary matrices. These unitaries describe the algorithmic freedom we have to process the unknown quantum state, and are determined by the target function $f(p)$. For each $\Omega_{i}$ we associate a unitary $U_{i}$ of the form

\begin{equation}
U_{i}=H(a_{i})=\left(\begin{array}{cc}\sqrt{1-a_{i}}&\sqrt{a_{i}}\\ \sqrt{a_{i}}&-\sqrt{1-a_{i}}\end{array}\right),
\end{equation}

which yields a coin that gives output one with probability

\begin{equation}
h_{a}(p)=\left(\sqrt{p (1-a)}-\sqrt{a(1-p)}\right)^{2}
\end{equation}

when measured in the computational basis.

We call the set of unitaries $G=\lbrace U_{i}\rbrace$ and so the quantum probability distribution is a function of both $p$ and $G$.

\begin{equation}
P_{p,G}\left(\prod_{i=1}^{|G|}V(\underline{v}_{i})\right)=\prod_{i=1}^{|G|}\prod_{j=1}^{|\underline{v}_{i}|}|\langle \underline{v}_{i,j}|U_{i}|p\rangle|^{2}.
\end{equation}

We are interested in functions where we can construct sets and probability measures such that there is an event that occurs with probability $f(p)$ and is discernible.

\emph{Definition 2:} A function $f:[0,1]\rightarrow [0,1]$ is \emph{q-constructible} if there exists an event $S$ and a probability measure $P_{p,G}$ such that $S$ is discernible with respect to $P_{p,G}$ and $P_{p,G}(S)=f(p)$

The following definitions will be used for Theorem 1.

\emph{Definition 3:} A function $f(p):[0,1]\rightarrow \lbrack 0,1]$ is simple and poly-bounded (SPB) iff it satisfies

\begin{enumerate}
\item $f$ is continuous.
\item Both $Z=\left\{ z_{i} : f\left(
z_{i}\right) =0 \right \}$  and $W=\left\{ w_{i} :f\left(
w_{i}\right) =1 \right \}$ are finite sets.
\item $\forall z\in Z$ there exists constants $c,\delta >0$ and and integer $k<\infty $ such that 
\begin{equation*}
c\left( p-z\right) ^{2k}\leq f\left( p\right) ~~\forall p\in \left[ z-\delta
,z+\delta \right]. 
\end{equation*}
\item  $\forall w\in W$ there exists constants $c,\delta >0$ and an integer $k<\infty $ such that 
\begin{equation*}
1-c\left( p-w\right) ^{2k}\geq f\left( p\right) ~~\forall p\in \left[
w-\delta ,w+\delta \right] .
\end{equation*}
\end{enumerate}

\emph{Definition 4:} A function  $L:[0,1]\rightarrow [0,1]$ is a lower bounding function for a function $f:[0,1]\rightarrow [0,1]$ if 
\begin{equation}
L(p) \leq f(p) \text{  for } p \in [0,1].
\end{equation}
A function  $U:[0,1]\rightarrow [0,1]$ is an upper bounding function for a function $f:[0,1]\rightarrow [0,1]$ if 
\begin{equation}
U(p) \geq f(p) \text{  for } p \in [0,1].
\end{equation}

We will need some more definitions for Theorem 2.

\emph{Definition 5:} A function $f:[0,1]\rightarrow [0,1]$ is finitely q-constructible if it is q-constructible and there exists some $N$ such that after observing the first $N$ bits of the string one can say with certainty whether the output will be $0$ or $1$.

\emph{Definition 6:} A function $\alpha : \Omega \rightarrow \mathbf{N} $ is \emph{discernible}, if $\alpha^{-1}(n)$ is discernible for every $n$

We now prove Proposition 1, which is the main result needed to adapt the classical result to the quantum case.

\emph{Proposition 1:}
For any SPB function $f(p)$ there exists a q-constructible pair of bounding functions $L(p)$ and 
$U(p)$ such that   
\begin{equation*}
L(p)\leq f\left( p\right) \leq U\left( p\right) 
\end{equation*}%
where  
\begin{equation*}
\max_{p\in \lbrack 0,1]}(U\left( p\right) -L\left( p\right)) <\frac{1}{2}.
\end{equation*}

These two conditions ensure that 
\begin{equation}
g(p)=\frac{L(p)}{1-U(p)+L(p)}
\end{equation}
will be a good approximation to $f(p)$, which will be an important step later on.

We will outline an explicit construction of these bounding functions, which
are key to Theorem 1. 

Consider some candidates for $L(p),U(p)$ which do not quite work, namely $%
A_{n}( p) ,B_{n}( p) $ defined by%
\begin{eqnarray*}
A_{n}( p)  &=&\sum_{k=0}^{n}\frac{2}{3}f\left( \frac{k}{n}\right)
{n \choose k}p^{k}( 1-p) ^{n-k} \\
B_{n}( p)  &=&\sum_{k=0}^{n}\left( \frac{1}{3}+\frac{2}{3}f\left( 
\frac{k}{n}\right) \right){n \choose k} p^{k}\left( 1-p\right) ^{n-k}
\end{eqnarray*}%

These bounding functions have a minimum separation of one third rather than one half. This is because some of the modifications we will use will slightly increase this value, meaning we require an initial buffer.

From Bernstein's proof of Weierstrass' approximation theorem[2] we know that $
A_{n}(p) ,B_{n}(p) $ converge \emph{uniformly} in $n$
to $\frac{2}{3}f(p) $ ,$\left( \frac{1}{3}+\frac{2}{3}f\left(
p\right) \right) ,$ and so would be good bounding functions had we the freedom to take $n$ to infinity. They are also manifestly constructible (within even a classical
Bernoulli factory). However in the vicinity of the zeros and ones of $%
f( p) $ we have a problem, as easily illustrated with the example[SupFig.1].

The strategy for dealing with this will be to multiply $A_{n}( p) $ by a generalized Heaviside-type
function (that must itself be q-constructible). We focus on $L\left( p\right) $ whenever it is obvious how to define the
corresponding procedure for $U\left( p\right) $. Defining 
\begin{equation}
h_{a}(p)=\left(\sqrt{p (1-a)}-\sqrt{a(1-p)}\right)^{2}
\end{equation}
consider the function%
\begin{equation*}
T_{m,M,z}\left( p\right) :=\left( 1-(1-h_{z}\left( p\right)) ^{m}\right)
^{M}
\end{equation*}%
which is plotted here for $M=3,z=1/2$ and $m=400$,[SupFig.2]:

Note that $T_{m,M,z}\left( z\right) =0$ and that $T_{m,M,z}\left(
p\right) $ is constructible with a finite number of quoins, since $%
h_{z}\left( p\right) $ is constructible with a single quoin upon which the
unitary transformation 
\begin{equation}
\left[ 
\begin{array}{cc}
\sqrt{1-z} & \sqrt{z} \\ 
-\sqrt{z} & \sqrt{1-z}%
\end{array}%
\right]
\end{equation}
is performed prior to a computational basis measurement.
Bernoulli's inequality[3] $\left( 1+x\right) ^{n}\geq 1+nx$ can be used
to show that%
\begin{eqnarray*}
\lim_{m\rightarrow \infty }T_{m,M,z}\left( p\right)  &=&0~~if~~h_{z}\left(
p\right) <1/M \\
&=&1~~if~~h_{z}\left( p\right) >1/M
\end{eqnarray*}%

We can use such functions to modify $A_{n}(p)$, to try and force it to always be a lower bound for $f(p)$. For the same example consider now the product $A_{100}(p)T_{400,5,1/2}(p)T_{400,1,3/4}(p)$:

There are a couple of issues we need to be careful about [SupFig.3]: 

Issue 1. When we \textquotedblleft push down\textquotedblright\ $A_{n}\left(
p\right) $ we increase its distance from $B_{n}\left( p\right) ,$ which
itself is going to be \textquotedblleft pushed up\textquotedblright\ at
other (possibly nearby) locations. So we run the risk of increasing the
distance between the lower- and upper-bounding functions to more than 1/2. You can see $A(p)$ being flattened too much around $p=0.5$

Issue 2. We want the generalized Heaviside function to go towards zero over a
finite width interval around $z$ that is not too narrow - it must include
all the problematic points at which $A_{n}(p)$ becomes larger than $f(p).$ You can see this happening around $p=0.75.$
When this happens we can either increase the interval width or
alternatively increase $n$ so that the points where $A_{n}(p)$ crosses $%
f\left( p\right) $ \textquotedblleft slide\textquotedblright\ inwards.

Since $h_{z}\left( z\right) =0$ and $h_{z}\left( p\right) $ is convex we
readily see that, denoting by $\mu _{i}$ the two solutions to $h_{z}\left(
p\right) =1/M,$ any large integer $M$ defines a finite interval $I(M,z):=%
\left[ \mu _{1},\mu _{2}\right] $ that (strictly) contains the point $z.$
The interval width decreases as $M$ is increased, but is finite for all $%
M<\infty $. Inside/outside $I(M,z)$ standard theorems for continuous
functions ensure that the convergence (in $m$) to $0$ or $1$
respectively is uniform.   

A series expansion%
\begin{equation*}
T_{m,M,z}\left( p\right) \approx \frac{\left( M(p-z)^{2}\right) ^{m}}{4z(1-z)%
}+O((p-z)^{4m})
\end{equation*}%
around the point $p=z$ makes clear the reason for the conditions 3
and 4 in Definition 1 - for functions that rise exponentially slowly from
any zero $z$ we would not be able to choose parameters $M$ and $m$ such that
we can push $A_{n}\left( p\right) $ below $f\left( p\right) $ in the
vicinity of $z$.

We now have all ingredients in place to outline a procedure for constructing
our lower-bounding function, which will take the form 
\begin{equation*}
L\left( p\right) =A_{n}\left( p\right) \prod_{i|z_{i}\in
Z}T_{m_{i},M_{i},z_{i}}\left( p\right) 
\end{equation*}%
for some appropriate choice of parameters $n,m_{i},M_{i}.$ For each zero $%
z_{i}\in Z$ we first choose large integers $M_{i}$ such that we fix
non-overlapping intervals $I\left( M_{i},z_{i}\right) $ which are smaller
than those determined by the associated $\delta $ (see Definition 1) for that 
$z_{i}.$ We ensure they also do not overlap with the similar intervals $%
I\left( M_{i},w_{i}\right) $ used in the construction of $U\left( p\right) .$
However these intervals remain finite so that as $n$ is increased, uniform convergence
ensures that all the problematic points where $A_{n}\left( p\right) \geq
f\left( p\right) $ end up strictly contained within the $I\left(
M_{i},z_{i}\right),$ so $L(p)\leq f(p)$.

Finally, and critically, we also ensure that 
\begin{equation*}
\sup_{I\left( M_{i},z_{i}\right) }\left( \frac{1}{3}+\frac{2}{3}f\left(
p\right) \right) <\frac{1}{2}~~~~\forall i
\end{equation*}%
This is possible, because $f$ is continuous and the interval contains a zero
of $f$ so for sufficiently large $n$ there will be an interval containing $z$ for which $U(p)<1/2$. The importance of this
condition is that now, even as our lower bounding function is taken close to
0 within the intervals $I\left( M_{i},z_{i}\right) ,$ our upper-bounding
function (which converges uniformly to $\left( \frac{1}{3}+\frac{2}{3}%
f\left( p\right) \right) $ within $I\left( M_{i},z_{i}\right) $) can be
brought within a distance $1/2$ from the lower bounding function.  

In summary, we now have that $L\left( p\right) $ converges uniformly on $%
[0,1]-\cup _{i}I\left( M_{i},z_{i}\right) $ to $\frac{1}{3}f\left( p\right) $
and on $\cup _{i}I\left( M_{i},z_{i}\right) $ to 0 as the remaining free
parameters $n,m_{i}$ are increased. Similarly we construct 
\begin{equation*}
U\left( p\right) =B_{n}\left( p\right) \prod_{i|w_{i}\in
W}T_{m_{i},M_{i},w_{i}}\left( p\right) +1-\prod_{i|w_{i}\in
W}T_{m_{i},M_{i},w_{i}}\left( p\right) 
\end{equation*}%
which converges uniformly on $[0,1]-\cup _{i}I\left( M_{i},w_{i}\right) $ to 
$\frac{1}{3}+\frac{2}{3}f\left( p\right) $ and on $\cup _{i}I\left(
M_{i},w_{i}\right) $ to 1. The conditions of Proposition 1 can clearly be
met with some suitable large (but finite) choice of the remaining free
parameters.

\emph{Theorem 1:} A function $f:[0,1]\rightarrow [0,1]$ is constructible with quoins and a finite set of single qubit unitaries if and only if it is SPB.

The proof of this theorem essentially follows that of Keane and
O'Brien[1].

\begin{proof}
Let $L_{k}\left( p\right) $ and $U_{k}\left( p\right) $ denote lower and
upper-bounding functions satisfying Proposition 1, for the sequence of SPB
functions  $f_{k}\left( p\right) ,$ where 
\begin{eqnarray*}
f_{1}\left( p\right)  &=&f\left( p\right)  \\
g_{k}\left( p\right)  &=&\frac{L_{k}\left( p\right) }{1-U_{k}\left( p\right)
+L_{k}\left( p\right) } \\
f_{k+1}(p) &=&\frac{4}{3}\left( f_{k}\left( p\right) -\frac{1}{4}g_{k}\left(
p\right) \right) 
\end{eqnarray*}%

Since $U(p)$ and $L(p)$ are finitely q-simulable there exists some $N$ after which we can say with certainty whether they occur or not. If neither occur, the output is heads; if both occur, the output is tails; if only $L(p)$ occurs, simulate the functions again.

This procedure produces an output with probability $(1-U(p)+L(p))$ on each trial and so the probabilty of this outputting heads is

\begin{equation}
g_{i}(p)=\frac{L(p)}{1-U(p)+L(p)}
\end{equation}

\begin{align}
f_{i+1}(p)&=\frac{4}{3}(f_{i}(p)-\frac{1}{4}.\frac{L(p)}{1-U(p)+L(p)})\\
&>\frac{4}{3}(f_{i}(p)-\frac{1}{2}.L(p))\\
&>\frac{4}{3}(f_{i}(p)-\frac{f_{a}(p)}{2})\\
&>\frac{2}{3}f_{i}(p)
\end{align}

Similarly

\begin{align}
f_{i+1}(p)&=\frac{4}{3}(1-(1-f_{i}(p))-\frac{1}{4}+\frac{1}{4}.\frac{1-U(p)}{1-U(p)+L(p)})\\
1-f_{i+1}(p)&=\frac{4}{3}((1-f_{i}(p))-\frac{1}{4}.\frac{1-U(p)}{1-U(p)+L(p)})\\
&<\frac{2}{3}(1-f_{i}(p))
\end{align}

So if $f_{i}(p)$ is SPB then so is $f_{i+1}(p)$. 

Since%
\begin{equation*}
\lim_{N\rightarrow \infty }\left[ f\left( p\right) -\sum_{k=0}^{N}\left( 
\frac{3}{4}\right) ^{k-1}\frac{1}{4}g_{k}\left( p\right) \right]
=\lim_{N\rightarrow \infty }\left( \frac{3}{4}\right) ^{N-1}f_{k}\left(
p\right) =0
\end{equation*}%
we have that $f\left( p\right) $ can be convexly decomposed%
\begin{equation*}
f=\sum_{k=0}^{\infty }\left( \frac{3}{4}\right) ^{k-1}\frac{1}{4}g_{k}\left(
p\right) .
\end{equation*}%
Using ancillary randomness to sample an index $k$ with probability $\left( 
\frac{3}{4}\right) ^{k-1}\frac{1}{4}$ we then construct a $g_{k}\left(
p\right) $ coin and we have shown that any SPB function can be constructed.

To show that any function constructible with quoins and a finite number of unitaries must be SPB is far simpler.

With the exception of the trivial functions $f(p)=0,1$, if a function is constructible then there is some number of flips $k$ such that after $k$ flips the probability of outputting 0 and the probability of outputting 1 are both greater than zero. 

The probability of outputting 1 must be less than $f(p)$ but be at least some constant multiplied by the least probable string which in turn must be greater than the expression below.

\begin{equation}
P(\text{output 1 }|k \text{ flips}) \geq\epsilon \prod_{i=1}^{n} (|\langle 0 |U_{i}|p\rangle \langle 1 |U_{i}|p\rangle|)^{4k}\geq \epsilon  \prod_{i|z_{i}\in
Z} (p-a_{i})^{k}
\end{equation}

This yields condition three of Definition 3 and half of condition two. Repeating the process with output one yields condition four and the other half of condition two. Finally we recall that a sum of continuous functions is a continuous function and since the probability measure on open sets give continuous functions and the event is made up of such sets the probability of the event must e a continuous function. Thus we have shown any function constructible with quoins and a finite number of unitaries must be SPB and we have completed the proof of Theorem 1.

\end{proof}

For the classical Bernoulli factory there are two different approachs which yield the same result. When we applied both approaches to the quantum cases we found that they led us in different directions. When emulating the approach due to Keane and O'Brien proof of the first theorem arose naturally, however when we tried to prove the first theorem with the approach due to W{\"a}stlund various complications arose. This led to us proving Theorem 2 instead.

\emph{Theorem 2:} A function $f:(0,a_{1})\cup(a_{1},a_{2})\cup ...\cup(a_{n},1)\rightarrow [0,1]$ is constructible with quoins and a finite set of single qubit unitaries if $f$ is continuous on its domain and there exists a finite list $\lbrace a_{1},a_{2},...a_{n'} \rbrace$, which contains $\lbrace a_{1},a_{2},...a_{n} \rbrace$, and integer $k$ such that 
\begin{equation}
a^{k}(p) \leq f(p) \leq 1-a^{k}(p)
\end{equation}
for all $p \in (0,1)$, where
\begin{equation}
a(p) := p(1-p)\prod_{1\leq i \leq n'} h_{a_{i}}(p)(1- h_{a_{i}}(p))
\end{equation}

This second theorem gives stranger functions than the first. At the set of excluded points $\lbrace a_{i} \rbrace$ the function can undergo a discontinuity or even have no definite value. Were the functions extended over the whole interval $(0,1)$ we would find we only require that the functions be piecewise continuous. The list $\lbrace a_{1},a_{2},...a_{n'} \rbrace$ contains all the points at which $f(p)$ is not defined and any points within the domain where $f(p)$ takes the value zero or one. We prove Theorem 2 by showing that we can pinpoint $p$ to within an interval with high confidence and then approximate $f(p)$; by averaging over many such approximations we achieve an exact sample. There are obvious parallels with promise problems.

The proof can be summarised as follows:

\begin{itemize}
  \item It is shown in proposition 2 that one can guess which of a set of intervals $p$ lies in with a small error which depends on $p$.
  \item It is shown in proposition 3 that we can choose our intervals such that there is a value $q_{i}$ which differs from the possible values of $f(p)$ in the interval by some small error which depends on $p$.
  \item It is shown in proposition 3 that these two results lead to an approximation of $f(p)$ which differs from the true function by a small error which depends on $p$
  \item As with Theorem 1, it is shown that if we can create good approximations we can create a sampling algorithm.
\end{itemize}

\emph{Proposition 2:} If $(0,a_{1})\cup(a_{1},a_{2})\cup ...\cup(a_{n},1)$ is covered by countably many open subintervals $W_{i}$, $i \in \mathbf{N}$, and $k$ is a positive integer, then there is an discernible function $\alpha : \Omega \rightarrow \mathbf{N} $ such that for every $p \in (0,1)$,
\begin{equation}
P_{p}(\alpha^{-1}\lbrace i :p \notin  W_{i} \rbrace ) < a^{k}(p)
\end{equation}

Proposition 2 states that given a set of intervals covering the domain of the function we can pinpoint $p$ to within one of them after a finite number of tosses with the probability that our guess is incorrect being less than $a^{k}(p)$. This is an essential step in our proof.

\bigskip

\noindent{\bf Proof}

In order to guess which interval we are in we must narrow down a potentially infinite number of intervals to a finite number. To this end we wish to restrict ourselves to a union of closed intervals so that we can invoke compactness.

For each $a \in\lbrace a_{1},a_{2},...a_{n} \rbrace$ we flip $h_{a_{i}}(p)$-coins until we have observed $nk+2k+1$ tails. In addition we flip $p$-coins until we have observed $nk+2k+1$ tails \emph{and} $nk+2k+1$ heads. For each case we call the number of flips before we see the required results $m_{a_{i}}$, $m_{0}$ for $p$-coins, and we call the maximum value $m_{max}$.

Now for some $a_{i}$ the probability of the required results occurring after $m_{a_{i}}$ is.

\begin{equation}
P(\geq (nk+2k+1) \text{ tails in } m_{a_{i}}\text{ flips}) < 2^{m_{a_{i}}}(1-h_{a_{i}}(p))^{nk+2k+1}
\end{equation}

We can use this to rule out a region around ${a_{i}}$ of size $2\epsilon$ as follows.

\begin{align}
P(\geq (nk+2k+1) \text{ tails in } m_{a_{i}}\text{ flips}|p\in\lbrace a_{i}-\epsilon,a_{i}+\epsilon\rbrace) &< 2^{m_{a_{i}}}(1-h_{a_{i}}(p))^{2nk+2k+1}\\
&<(1-h_{a_{i}}(p))^{2nk+2k}\\
&<\epsilon^{2nk+2k}\\
&< a^{k}(p)
\end{align}

The same procedure can also show the probability that $p \in (0,\epsilon]$ or $p \in [1-\epsilon,1)$ is also below $a^{k}(p)$. This means we can say that $p \in [\epsilon,a_{1}-\epsilon]\cup [a_{1}+\epsilon,a_{2}-\epsilon]\cup ...\cup [a_{n}+\epsilon,1-\epsilon]$ with probability greater than $1-a^{k}(p).$ It is possible for some of these regions to be large enough that they merge; this is not a problem.

Now that we are restricted to a union of closed intervals we can, by compactness, choose a finite number of $W_{i}$'s which cover this set. We can also find $\delta>0$ such that $a^{k}(p)\geq \delta$ throughout this region. From the finite open cover we can choose a closed interval $F_{i}\subset W_{i}$ for each $i$ such that the $\lbrace F_{i} \rbrace$ cover our union of closed intervals.   Now we estimate $p$ by flipping classical coins and calculating the frequency of heads, our estimated value is

\begin{equation}
f_{N}=\frac{X_{1}+...+X_{N}}{N}.
\end{equation}

We do this with $N$ flips, $N$ being chosen such that

\begin{equation}
P(f_{N}\in F_{i}|p\notin W_{i})<\delta
\end{equation}

We now define our discernible function $\alpha$ as $\alpha(f_{N} \in F_{i})=i$. Recall that the $\lbrace F_{i}\rbrace$ are conditioned on $m_{max}$ so the function is not as simple as it appears.

We have now shown that such a discernible function exists.

$\qed$

\emph{Proposition 3:} If f is any continuous function $(0,a_{1})\cup (a_{1},a_{2})...  \cup (a_{n},1) \rightarrow (0,1)$, and l is any positive integer, then there is a function g representing the probability of an discernible event such that for every p in $(0,a_{1})\cup (a_{1},a_{2})...  \cup (a_{n},1)$,

\begin{equation}
|f(p)-g(p)|<a^{l}(p)
\end{equation}

Proposition 3 uses Proposition 2 to show that we can find contructible functions which are good approximations to $f(p)$ which allows us to complete our proof of Theorem 2 in a manner similar to that for Theorem 1; we sample a randomly from a set of approximations to $f(p)$ in such a way that on average we sample $f(p)$.

\noindent{\bf Proof}

Since $f$ is continuous, we can choose countably many subintervals $W_{1},W_{2},W_{3},...$ covering $(0,a_{1})\cup (a_{1},a_{2})...  \cup (a_{n},1)$, and numbers $q_{1},q_{2},q_{3},...$ so that for every $p \in W_{i}$

\begin{equation}
|f(p)-q_{i}|<a^{l+1}(p).
\end{equation}

We have shown in Proposition 2 that we can guess the interval with some small error probability. The above statements show that we can choose our intervals such that they have the $q$ values we need. The idea is that we guess our interval, $W_{i}$, and that we have chosen our intervals such that $f(p)$ can be approximated as $q_{i}$ within them.

In the event that $f$ takes the value zero or one at a point within its domain then we assign a function $q_{i}(p)$ to an interval containing that point such that for every $p \in W_{i}$

\begin{equation}
|f(p)-q_{i}(p)|<a^{l+1}(p)
\end{equation}
using the techniques from Theorem 1. For ease of notation we will continue the proof assuming this is not necessary

By Proposition 2, there is an discernible function $\alpha : \Omega \rightarrow \mathbf{N} $  such that for every $p$,

\begin{equation}
P_{p}(\alpha^{-1}\lbrace n :p \notin W_{n} \rbrace ) < a^{l+1}(p)
\end{equation}

Once $\alpha$ is known, which we know will be in finite time, we then use an event which occurs with probability $q_{\alpha}$.

Thus the total event has probability

\begin{equation}
g(p)=\sum_{i=1}^{\infty}P_{p}(\alpha^{-1}(i))q_{i}
\end{equation}

Hence for every $p$, the error $|f(p)-g(p)|$ is bounded by

\begin{equation}
P_{p}(p\notin U_{\alpha} )+\text{sup}\lbrace|f(p)-q_{i}|:p \in W_{i}\rbrace=2a^{l+1}(p),
\end{equation}

the chance of guessing the wrong interval plus the difference between $q_{i}$ and $f(p)$.

Recalling that

\begin{equation}
a(p) := p(1-p)\prod_{1\leq i \leq n'} h_{a_{i}}(p)(1- h_{a_{i}}(p))\leq p(1-p) \leq \frac{1}{2},
\end{equation}

and therefore

\begin{equation}
 P_{p}(p\notin W_{\alpha} )+\text{sup}\lbrace|f(p)-q_{i}|:p \in U_{i}\rbrace=2a^{l+1}(p)< a^{l}(p).
\end{equation}

\noindent So we have shown the total error is less than the error we are allowed. $\qed$

We can now complete the proof Theorem 2 by showing that this ability to create approximations allows us to create an algorithm made from successive approximations.

\noindent{\bf Proof}

Let  $f:(0,a_{1})\cup(a_{1},a_{2})\cup ...\cup(a_{n},1)\rightarrow [0,1]$ be any continuous function satisfying

\begin{equation}
a^{k}(p)<f(p)<1-a^{k}(p)
\end{equation}
for some $k$. If we apply Proposition 3 to $f$ with $l=k+1$ we find that there exists a function $f_{1}$, representing the probability of an discernible event, such that, for every $p$,

\begin{equation}
|f_{1}(p)-f(p)|<a^{k+1}(p).
\end{equation}
We then have
\begin{equation}
f_{1}(p)>f(p)-a^{k+1}(p).
\end{equation}
Since $f(p)<1-a^{k}(p)$, we see that
\begin{equation}
f_{1}(p)>2f(p)-a^{k+1}(p)-1+a^{k}(p)>2f(p)-1+a^{k+1}(p).
\end{equation}
Similarly,
\begin{equation}
f_{1}(p)<f(p)+a^{k+1}(p),
\end{equation}
Since $f(p)>a^{k}(p)$, we see that
\begin{equation}
f_{1}(p)<2f(p)+a^{k+1}(p)-a^{k}(p)<2f(p)-a^{k+1}(p).
\end{equation}
Combining these gives
\begin{equation}
2f(p)-a^{k+1}(p)<f(p)<2f(p)-1+a^{k+1}(p)
\end{equation}
which can be rearranged as
\begin{equation}
a^{k+1}(p)<2f(p)-f_{1}(p)<1-a^{k+1}(p).
\end{equation}

If we repeat this argument, replacing $f(p)$ by $2f(p)-f_{1}(p)$ and $k$ by $k+1$, we find that these exists a function $f_{2}$ which is the probability of an discernible event.

\begin{equation}
a^{k+2}(p)<4f(p)-2f_{1}(p)-f_{2}(p)<1-a^{k+2}(p)
\end{equation}

Continuing in this way we can find functions $f_{1},f_{2},f_{3},...$ each representing the probability of an discernible event, such that for every $n$,

\begin{equation}
a^{k+n}(p)<2^{n}f(p)-2^{n-1}f_{1}(p)-...-f_{n}(p)<1-a^{k+n}(p)
\end{equation}

If we replace the lower and upper bounds by $0$ and $1$ and divide  by $2^{n}$ we get

\begin{equation}
0<f(p)-\frac{1}{2}f_{1}(p)-...-\frac{1}{2^{n}}f_{n}(p)<\frac{1}{2^{n}})
\end{equation}

Letting $n \rightarrow \infty$, we obtain
\begin{equation}
f(p)=\sum_{n=1}^{\infty}\frac{1}{2^{n}}f_{n}(p)
\end{equation}
So if we choose a value $n$ with probability $p_{n}=\frac{1}{2^{n}}$ and then sample $f_{n}(p)$ we have sampled $f(p)$ exactly. 

To show that any function constructible with quoins and a finite number of unitaries must obey the conditions of Theorem 2 is far simpler.

With the exception of the trivial functions $f(p)=0,1$, if a function is constructible then there is some number of flips $k$ such that after $k$ flips the probability of outputting 0 and the probability of outputting 1 are both greater than zero. 

The probability of outputting 1 must be less than $f(p)$ but be at least some constant multiplied by the least probable string which in turn must be greater than the expression below.

\begin{equation}
P(\text{output 1 }|k \text{ flips}) >\epsilon \prod_{i=1}^{n} (|\langle 0 |U_{i}|p\rangle \langle 1 |U_{i}|p\rangle|)^{4k}>a^{k}(p)
\end{equation}

The least probable string after k flips occurs with probability  $\text{min}_{a_{i}}(1-h_{a_{i}}(p))^{k}$

\begin{equation}
\text{min}_{a_{i}}(1-h_{a_{i}}(p))^{k} > \left( p(1-p)\prod_{1\leq i \leq n'} h_{a_{i}}(p)(1- h_{a_{i}}(p))\right)^{k}=a^{k}(p)
\end{equation}

Repeating the process with output zero and combining the two yields.

\begin{equation}
a^{k}(p) \leq f(p) \leq 1-a^{k}(p),
\end{equation}
as required.

 Finally we recall that a sum of continuous functions is a continuous function and since the probability measure on open sets give continuous functions and the event is made up of such sets the probability of the event must be a continuous function. It should be emphasised that although the extensions of many of these functions are piecewise continuous the functions themselves are always continuous on their domain. This completes the proof of Theorem 2.$\qed$

\newpage

\section*{Supplementary References}

\begin{enumerate}

\item  M. S. Keane and G. L. O'Brien. A Bernoulli factory. ACM Trans. Model. Comput. Simul. 4(2):213-219, April 1994

\item W. Rudin. Principles of mathematical analysis, vol. 3. McGraw-Hill New York, 1976

\item N. L. Carothers. Real analysis. Cambridge University Press, 2000.

\end{enumerate}

\end{document}